\documentclass{article}

\usepackage{spconf,amsmath,graphicx,hyperref}
\usepackage{subfig} 
\usepackage{xcolor}
\usepackage{fontawesome} 
\usepackage{pifont}
\usepackage{natbib}
\usepackage{multirow}
\usepackage{booktabs}
\usepackage{ragged2e}
\title{Fake Speech Wild: Detecting Deepfake Speech on Social Media Platform}
\name{
	Yuankun Xie$^{1}$, Ruibo Fu$^{2}$\textsuperscript{*}, 
	Xiaopeng Wang$^{3}$, Zhiyong Wang$^{2}$, Ya Li$^{5}$, Zhengqi Wen$^{4}$, \\ 
	 Haonnan Cheng$^{1}$, Long Ye$^{1}$\textsuperscript{*}\thanks{\textsuperscript{*}Corresponding authors.}
}

\address{
	$^{1}$Communication University of China, Beijing, China\\
	$^{2}$Institute of Automation, Chinese Academy of Sciences, Beijing, China\\
	$^{3}$Beijing Institute of Technology, Beijing, China\\
	$^{4}$Beijing National Research Center for Information Science and Technology, Tsinghua University\\
	$^{5}$Beijing University of Posts and Telecommunications, Beijing, China
}

\begin{document}
\ninept
\maketitle

\begin{abstract}
	The rapid advancement of speech generation technology has led to the widespread proliferation of deepfake speech across social media platforms. While deepfake audio countermeasures (CMs) achieve promising results on public datasets, their performance degrades significantly in cross-domain scenarios. To advance CMs for real-world deepfake detection, we first propose the Fake Speech Wild (FSW) dataset, which includes 254 hours of real and deepfake audio from four different media platforms, focusing on social media. As CMs, we establish a benchmark using public datasets and advanced self-supervised learning (SSL)-based CMs to evaluate current CMs in real-world scenarios. We also assess the effectiveness of data augmentation strategies in enhancing CM robustness for detecting deepfake speech on social media. Finally, by augmenting public datasets and incorporating the FSW training set, we significantly advanced real-world deepfake audio detection performance, achieving an average equal error rate (EER) of 3.54\% across all evaluation sets.  The FSW dataset is publicly available\footnote{ \url{https://github.com/xieyuankun/FSW}}.
\end{abstract}

\section{Introduction}
With the current development of speech generation technologies such as Text-to-Speech (TTS) and Voice Conversion (VC), synthesized voices are becoming increasingly realistic, which we categorized as deepfake audio.
The recent emergence of audio language model (ALM) \cite{chu2024qwen2, ju2024naturalspeech, chen2025neural, du2025cosyvoice} has made the production process of deepfake audio more convenient, allowing anyone to easily synthesize such deepfake audio and upload it to social media platforms. Although deepfake technology significantly reduces production costs for creators, it can potentially be used to impersonate speakers, create false information, and even generate negative public opinion, posing a significant challenge for regulators and society as a whole.

In response to the increasing emergence of deepfake audio, Audio Deepfake Detection (ADD) technology has been developed. Various initiatives and competitions, such as ASVspoof \cite{todisco19_interspeech, liu2023asvspoof, wang2024asvspoof5} and the Audio Deepfake Detection Challenge \cite{yi2022add, yi2023add}, have been established to promote research focused on developing countermeasure (CM) against deepfakes. Currently, state-of-the-art (SOTA) CMs typically achieve an Equal Error Rate (EER) of less than 1\% on public datasets \cite{zhang2024improving, wang24ga_interspeech, wang24ca_interspeech, wang2024mixture, truong24b_interspeech}. However, they suffer from performance declines when dealing with cross-domain scenarios.
\begin{figure*}[!t]
	\centering
	\subfloat{\includegraphics[width=6in]{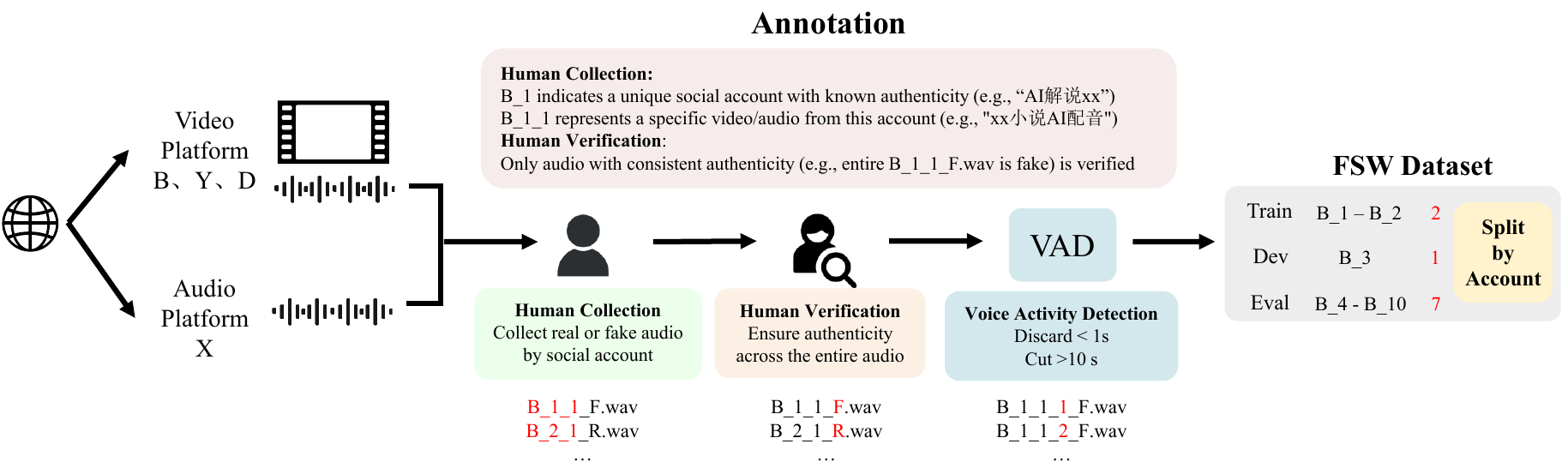}}
	\hfil
	\caption{The construction pipeline of FSW dataset.}
	\label{fig:dataset} 
	\vspace{-10pt}
\end{figure*}
Muller et al. were the first to propose this generalization issue \cite{muller22_interspeech}. This work collected genuine and deepfake audio of well-known political speakers from the social media platform YouTube, established an English dataset called "In the Wild" (ITW), and trained several CMs to test on the ITW dataset. Surprisingly, the CMs trained on the ASVspoof2019LA (19LA) \cite{todisco19_interspeech} training set achieved an average Equal Error Rate (EER) of 9.85\% on the 19LA test set, but the EER exceeded 60\% on the ITW dataset. This work is the first to reveal the generalization issue in the field of ADD.   

Although ITW is the first to explore the issue of CM generalization in detecting deepfake speech on social media platforms, we believe there are still areas that warrant further optimization. From the perspective of the dataset, ITW is limited to a single platform (YouTube) and a single language (English). Additionally, the speech deepfake methods used in ITW are relatively early, and at that time, there was not a significant amount of ALM-based deepfake audio. This raises uncertainty about whether the CMs that perform well on ITW will maintain their effectiveness across other languages, social media platforms, or novel deepfake methods. 

From the perspective of CMs, the effectiveness of current CMs on public datasets can be attributed to the "cleanliness" of the data. Real audio samples are typically recorded in controlled studio environments, such as those used in the VCTK \cite{Veaux2017CSTRVC} and ALSHELL \cite{shi21c_interspeech} datasets. Similarly, deepfake audio samples are often synthesized using models trained on these high-quality real audio recordings. 
However, the variety of speech scenarios on modern social media platforms, such as audiobooks, interviews, and news, which results in a domain discrepancy for existing CMs. This discrepancy mainly arises from differences in audio content types, recording environments and compression encodings used by different media platforms. It is necessary to explore how to design generalized CMs to address this domain discrepancy.


In this paper, we explore CMs for deepfake speech on Chinese social media platforms. We first constructed the Fake Speech Wild (FSW) dataset, which includes speech samples from 128 social accounts collected from four well-known social media websites. Since social accounts typically post consistent content, such as using the same voice for real or deepfake speakers, we can ensure the authenticity of each audio and video through initial social account screening combined with manual verification. Voice Activity Detection (VAD) was employed to segment each audio sample, which led to the division into 146,097 separate audio clips, amounting to 254.58 hours of audio data in total. Regarding the CMs, we first established a benchmark using existing public datasets and self supervised learning (SSL)-based CMs. Evaluations were then conducted on the evaluation sets of the public datasets as well as social media evaluation sets (ITW and FSW). We also investigated the effectiveness of different noise data augmentation (DA) techniques in mitigating the effects of recording environments and the diverse compression encodings employed by various media platforms. Finally, joint training was performed by combining the augmented public datasets with the FSW training set. Experimental results demonstrate that the optimal CM achieved an average EER across all evaluation sets of 3.54\%.

\vspace{-3pt}
\section{FSW Dataset Construction}
\vspace{-3pt}
In this section, we introduce the construction process of the FSW dataset, which includes the collection of FSW data samples, data processing, and the division process.

\vspace{-3pt}
\subsection{Human Collection}
Firstly, we selected four popular social media platforms for Chinese speech collection: Bilibili, YouTube, Douyin, and Ximalaya, hereafter referred to as B, Y, D, and X, respectively. Among these, B, Y, and D are video platforms, while X is an audio platform. We collected videos or audio files based on social media accounts, as the content posted by each account generally has the same authenticity attributes. This indicates that the speech content is often created by the same real speaker or through AI dubbing. Consequently, the number of accounts on short video or audio platforms is greater than on long video platforms. As indicated in Table \ref{tab:fsw}, the number of D and X accounts is much higher compared to B and Y. After the collection, we extracted the audio tracks from the videos and manually checked each collected audio file to confirm the authenticity of the entire audio. 

As shown in Figure \ref{fig:dataset}, we also adopt a standardized naming convention for the collected audio in this stage. The first field indicates the platform source, the second represents the account ID, the third denotes the audio index under that account, and the fourth specifies the authenticity label. This label is initially determined by the collector based on the account name, the title of the video or audio, as most of the collected samples explicitly indicate the use of AI synthesis in either the title or the account name, and the perceptual judgment of the specific audio content. 
\begin{table}[t]
	\vspace{-5pt}
	\centering
	\caption{Statistics of datasets in terms of accounts, segments, and duration (hours).}
	\label{tab:fsw}
	\fontsize{7.5pt}{8pt}\selectfont
	\setlength{\tabcolsep}{2pt} 
	\begin{tabular}{cccc|ccc|ccc}
		\toprule
		\multirow{2}{*}{Platform} & \multicolumn{3}{c|}{Account} & \multicolumn{3}{c|}{Segment} & \multicolumn{3}{c}{Duration (h)} \\
		\cmidrule(lr){2-4} \cmidrule(lr){5-7} \cmidrule(lr){8-10}
		& Real & Fake & Total & Real & Fake & Total & Real & Fake & Total \\
		\midrule
		B & 9 & 9 & 18 & 13,563 & 29,691 & 43,254 & 20.29 & 41.49 & 61.78 \\
		Y & 7 & 8 & 15 & 13,163 & 14,953 & 28,116 & 23.38 & 34.93 & 58.31 \\
		D & 11 & 7 & 18 & 14,643 & 14,983 & 29,626 & 32.54 & 31.74 & 64.28 \\
		X & 55 & 22 & 77 & 29,508 & 15,593 & 45,101 & 41.48 & 28.73 & 70.21 \\
		\midrule
		Total & 82 & 46 & 128 & 70,877 & 75,220 & 146,097 & 117.69 & 136.89 & 254.58 \\
		\bottomrule
	\end{tabular}

\end{table}

\begin{table*}[t]
	
	\centering
	\caption{EER (\%) ↓ results for CMs trained on public datasets. The co-trained dataset includes 19LA, CFAD, and Codecfake training sets.}
	\label{tab:publicdataset}
	\fontsize{7.5pt}{8pt}\selectfont
	\vspace{-10pt}
	\begin{tabular}{p{2cm}p{2cm}cccccccccc}
		\toprule
		\multirow{2}[4]{*}{Training set} & \multirow{2}[4]{*}{Countermeasure} & \multicolumn{4}{c}{Public Dataset} & \multirow{2}[4]{*}{FSW} & \multirow{2}[4]{*}{AVG} & \multicolumn{4}{c}{FSW Dataset} \\
		\cmidrule(lr){3-6} \cmidrule(lr){9-12}
		& & 19LA & CFAD & Codecfake & ITW & & & FSW\_B & FSW\_Y & FSW\_D & FSW\_X \\
		\midrule
		19LA & AASIST & 0.82 & 47.12 & 11.97 & 43.5 & 46.11 & 29.90 & 49.29 & 46.35 & 42.99 & 35.93 \\
		19LA & WavLM-AASIST & 0.46 & 15.37 & 0.23 &\bf 13.21 & 36.01 & 13.06 & 40.82 & 46.85 &\bf 25.14 & 33.06 \\
		19LA & XLSR-AASIST & \textbf{0.22} & \bf 8.81 &\bf 0.16 & 13.58 & \bf 32.51 &\bf 11.06 &\bf 30.90 &\bf 30.46 & 26.21 &\bf 31.54 \\
		\midrule
		CFAD & AASIST & 39.45 & 0.86 & 36.40 & 47.12 & 45.43 & 33.85 & 46.15 & 38.37 & 31.50 & 46.88 \\
		CFAD & WavLM-AASIST &\bf 3.51 & \textbf{0.08} & 0.74 &\bf 13.78 & 39.81 & 11.58 & 45.51 &\bf 22.86 & 27.13 & 43.27 \\
		CFAD & XLSR-AASIST & 5.03 & 0.71 &\bf 0.54 & 15.53 &\bf 33.62 &\bf 11.09 &\bf 41.18 & 24.70 & \textbf{21.33} &\bf 32.11 \\
		\midrule
		Codecfake & AASIST & 35.90 & 33.64 & \textbf{0.09} & 41.91 & 42.14 & 30.74 & 31.41 & 49.07 & 30.05 & 38.93 \\
		Codecfake & WavLM-AASIST & 12.29 & 14.59 & 0.18 & 27.05 & 34.54 & 17.73 & 23.05 & 26.26 & 34.60 & 41.04 \\
		Codecfake & XLSR-AASIST &\bf 4.54	&\bf 4.94	&0.16	&\bf 11.77	&\bf 21.82	&\bf 8.65
		&\bf 15.44	&\bf 13.47	&\bf 27.21&\bf19.15\\
		\midrule
		Co-trained  & AASIST & 7.36 & 2.67 &\bf 0.14 & 13.90 & 39.27 & 12.67 & 37.10 & 47.57 &\bf 21.78 & 40.82 \\
		Co-trained  & WavLM-AASIST & 0.73 & 2.97 & 0.34 & 18.88 & 31.46 & 10.88 & 36.41 & 22.13 & 26.60 & 32.20 \\
		Co-trained & XLSR-AASIST &\bf 0.54 &\bf 0.21 &\bf 0.14 & \textbf{9.57} & \textbf{17.23} & \textbf{5.54} & \textbf{12.93} & \textbf{11.53} & 23.13 & \textbf{15.29} \\
		\bottomrule
	\end{tabular}
	\vspace{-10pt}
\end{table*}
\subsection{Expert Verification}
After the collection process, four human expert ensures that each audio sample collected from the four platforms has consistent global authenticity. Specifically, if a real-labeled audio contains any fake segments, or a fake-labeled audio contains any real segments, the sample is considered invalid and will be discarded.
\vspace{-3pt}
\subsection{Voice Activity Detection}
After human experts confirmed the authenticity of each audio sample, we utilized pyannote\footnote{https://huggingface.co/pyannote/segmentation} for VAD segmentation of the collected audio. During segmentation, we discarded any segments shorter than 1 second. Since many platforms' real or AI-generated voiceovers often remove silent sections, leading to excessively long speech segments, we applied a rule to further segment any audio longer than 10 seconds into 10-second intervals until the segments were within 10 seconds. For the segmented clips derived from the original video or audio, the fourth field in the filename reflects their sequence in chronological order. As a result, we obtained 70,877 genuine segments and 75,220 fake segments, totaling 146,097 audio samples and 254.58 hours in duration. Dataset statistics after VAD are shown in Table \ref{tab:fsw}.
\vspace{-10pt}
\subsection{FSW Dataset Construction}
After VAD process, we constructed the FSW dataset by account into training, development, and evaluation sets in a 2:1:7 ratio. Due to the abundance of short videos or audio on platforms D and X, more accounts were included to ensure a balanced duration across all platforms. Moreover, the accounts used in the training, validation, and test sets are non-overlapping, which guarantees a fair evaluation of generalization ability.

The number of segments in terms of FSW dataset shown in Table \ref{tab:fsw_statistics}. We allocated 70\% of FSW dataset to the test set to ensure that the FSW dataset primarily as a test dataset for evaluating CM performance in real-world social media environments. On the other hand, we designated a small portion of the training and validation sets to evaluate whether a limited dataset containing only a few deepfake methods from real-world data can learn domain-invariant information beyond authenticity attributes—such as website-specific encoding and decoding patterns—to improve the robustness of the CM.

\vspace{-5pt}
\begin{table}[t]
	\centering
	\caption{Subset Statistics of FSW Dataset.}
	\label{tab:fsw_statistics}
	\vspace{-10pt}
	\fontsize{7.5pt}{8pt}\selectfont
	\begin{tabular}{c|ccc}
		\toprule
		FSW & Real & Fake & Total \\
		\midrule
		Train & 11,559 & 8,403 & 19,962 \\
		Dev & 5,214 & 12,718 & 17,932 \\
		Eval & 54,104 & 54,099 & 108,203 \\
		\midrule
		ALL & 70,877 & 75,220 & 146,097 \\
		\bottomrule
	\end{tabular}
	\vspace{-10pt}
\end{table}
\vspace{-7pt}
\section{Countermeasure}
\vspace{-5pt}
In this section, we delineate the construction of countermeasures by presenting the public training and testing data, baseline models, and DA methods.

\vspace{-7pt}
\subsection{Public Dataset}
\vspace{-5pt}
\label{section:public}
We first investigated the CM performance trained with the public datasets and tested on their corresponding test sets, ITW, and FSW. The information for all datasets involved in this work is shown below:

\textbf{ASVspoof2019LA (19LA)}: The 19LA training set comprises 25,380 audio samples and includes six spoofing methods (A01-A06), while the test set consists of 71,237 audio samples and includes thirteen spoofing methods (A07-A19). 19LA is a benchmark dataset for ADD research. Adopting the 19LA protocol for training and testing provides a reliable basis for evaluating the performance of CMs.

\textbf{Codecfake \cite{xie2025codecfake}}: Given the current popularity of ALM generated deepfake audio, social media platforms may contain a large amount of ALM-based deepfake audio. Therefore, we selected the Codecfake dataset, which is designed for ALM-based deepfake audio detection. The training set comprises 740,747 audio samples, generated from the VCTK and AISHELL source domains using six neural codecs. To evaluate the effectiveness of ALM-based audio detection, we conducted tests using the Codecfake A1 condition generated by VALL-E \cite{wang2023neural}.

\textbf{CFAD \cite{ma2024cfad}}: Considering the language of FSW and the codec issues across various platforms, we selected CFAD, the most comprehensive Mandarin ADD dataset, comprising genuine samples from six different domains and synthesized using twelve distinct methods. To enhance the CM performance on FSW, we utilized the codec-version of the CFAD dataset for training, validation, and testing. This version contains audio processed with various traditional codec methods, such as MP3, OGG, m4a, etc. For the test set, we selected the test seen subset for evaluation.

\textbf{In the wild (ITW) \cite{muller22_interspeech}}: ITW dataset collects both real and fake audio samples from publicly available sources like social networks and video streaming platforms. It potentially contains background noises and is designed to assess the generalizability of ADD models. 

\subsection{Data Augmentation}
\vspace{-5pt}
In order to tackle the robustness challenges posed by the complex audio environment on social media websites and the effects of codecs, we considered the following two strategies for augmenting the public training dataset with noise:

\textbf{MUSAN \& RIR (MR) \cite{David2015MUSAN, Tom2017A}}: Considering the abundance of audio with background music (BGM) and the different recording environments in FSW, we utilize the MR strategy for noise augmentation, a frequently used DA technique in the areas of speech recognition and speaker identification. Specifically, an online DA strategy was implemented with randomized selection among the following configurations: (1) no augmentation, (2) MUSAN speech subset, (3) MUSAN noise subset, (4) MUSAN music subset, (5) RIR dataset.

\textbf{Rawboost (RB) \cite{tak2022rawboost}}: RB is a signal based DA method applied directly to raw waveforms, has demonstrated promising performance improvements on both the telephony channel-based ASVspoof 2021 LA dataset and the compression codec-generated ASVspoof 2021 DF dataset. To investigate whether RB-enhanced CMs can maintain robustness against deepfake audio across diverse real-world platforms, we examine four distinct configurations: (1) RB1: Linear and non-linear convolutive noise, (2) RB2: Impulsive signal-dependent additive noise, (3) RB3: Stationary signal-independent additive noise, (4) RB4: Sequential application of (1)+(2)+(3).

\begin{table*}[t]
	\centering
	\caption{EER (\%) ↓ results for XLSR-AASIST trained on co-trained dataset using different data augmentation strategies.}
	\label{tab:da}
	\fontsize{7.5pt}{8pt}\selectfont
	\vspace{-10pt}
	\begin{tabular}{p{2cm}p{2.5cm}cccccccccc}
		\toprule
		\multirow{2}[4]{*}{Training set} & \multirow{2}[4]{*}{Data Augmentation} & \multicolumn{4}{c}{Public Dataset} & \multirow{2}[4]{*}{FSW} & \multirow{2}[4]{*}{AVG} & \multicolumn{4}{c}{FSW Condition} \\
		\cmidrule(lr){3-6} \cmidrule(lr){9-12}
		& & 19LA & CFAD & Codecfake & ITW & & & FSW\_B & FSW\_Y & FSW\_D & FSW\_X \\
		\midrule
		Co-trained & MR & \textbf{0.43} & 0.31 & 0.20 & 3.58 & 19.08 & \textbf{4.72} & 18.58 & 14.86 & 20.08 & \bf 12.67 \\
		Co-trained& RB1 & 0.45 & 0.74 & \textbf{0.14} & 5.26 & 20.09 & 5.34 & 20.59 & 15.07 & 22.73 & 17.97 \\
		Co-trained & RB2 & 0.45 & \textbf{0.25} & \textbf{0.14} & 8.79 & 19.47 & 5.82 & 19.03 & 14.73 & \textbf{19.01} & 16.47 \\
		Co-trained & RB3 & 0.50 & 0.44 & 0.20 & 6.03 & \bf 18.80 & 5.19 & \textbf{17.94} & 13.32 & 21.18 & 12.81 \\
		Co-trained & RB4 & 1.32 & 1.36 & 0.27 & \textbf{2.42} & 23.39 & 5.75 & 26.65 & \bf 11.48 & 21.67 & 21.46 \\
		Co-trained & RB3 + MR & 1.74 & 1.74 & 0.18 & 3.46 & 25.33 & 6.49 & 28.99 & 15.08 & 20.19 & 26.71 \\
		Co-trained  & RB4 + MR & 2.22 & 2.54 & 0.16 & 2.84 & 27.36 & 7.02 & 33.38 & 14.35 & 20.97 & 26.94 \\
		\bottomrule
	\end{tabular}
\end{table*}

\begin{table*}[t]
	\vspace{-5pt}
	\centering
	\caption{EER (\%) ↓ results for countermeasures trained on FSW training set and Co-trained dataset.}
	\label{tab:four_training}
	\fontsize{7.5pt}{8pt}\selectfont
	\vspace{-10pt}
	\begin{tabular}{p{2.6cm}p{1.9cm}cccccccccc}
		\toprule
		\multirow{2}[4]{*}{Training set} & \multirow{2}[4]{*}{Countermeasure} & \multicolumn{4}{c}{Public Dataset} & \multirow{2}[4]{*}{FSW} & \multirow{2}[4]{*}{AVG} & \multicolumn{4}{c}{FSW Condition} \\
		\cmidrule(lr){3-6} \cmidrule(lr){9-12}
		& & 19LA & CFAD & Codecfake & ITW & & & FSW\_B & FSW\_Y & FSW\_D & FSW\_X \\
		\midrule
		FSW & AASIST & \bf 39.12 & 42.84 & \textbf{34.42} & 47.94 & 20.35 & 36.93 & 19.10 & 14.25 & 30.88 & 20.46 \\
		FSW & WavLM-AASIST & 47.90 & 40.92 & 35.14 & \bf 39.90 & 19.71 & \bf 36.71 & \textbf{17.03} & \textbf{13.51} & 45.18 & 17.17 \\
		FSW & XLSR-AASIST & 48.93 & \textbf{39.51} & 37.54 & 48.05 & \textbf{16.40} & 38.09 & 17.50 & 14.30 & \textbf{25.17} & \textbf{13.17} \\
		\midrule
		Co-trained + FSW & AASIST & 13.30  & 3.00     & 0.34  & 33.23 & 29.63 & 15.90  & 26.69 & 30.49 & 41.36 & 25.00 \\
		Co-trained + FSW & WavLM-AASIST & 2.53  & 2.29  & \textbf{0.05} & 19.19 & 17.31 & 8.27  & 14.12 & \textbf{11.23} & 29.09 & 16.30 \\
		Co-trained + FSW & XLSR-AASIST  & 0.57  & \bf 0.13  & 0.23  & 9.35  & 12.55 & 4.57  & \textbf{9.71} & 12.16 & 22.05 & 9.99 \\
		Co-trained (MR) + FSW & XLSR-AASIST & \bf 0.45  & 0.21  & 0.20   & \bf 5.24  & \textbf{11.58} & \textbf{3.54} & 13.21 & 13.03 & \textbf{16.67} & \textbf{6.62} \\
		\bottomrule
	\end{tabular}
	\vspace{-10pt}
\end{table*}
\vspace{-8pt}
\subsection{Baseline Models}
This section introduces the baseline models used for ADD, including three representative models in the field of ADD: AASIST \cite{jung2022aasist}, WavLM-AASIST and XLSR-AASIST \cite{tak2022automatic}. For the backbone network, the state-of-the-art (SOTA) backend classification model AASIST was selected, which employs graph-based attention mechanisms to simultaneously capture spectral and temporal feature of audio signals. For the front-end, we explored raw audio waveforms, frozen self-supervised feature WavLM-large\footnote{https://huggingface.co/microsoft/wavlm-large} \cite{chen2022wavlm} with AASIST (WAVLM-AASIST), and Wav2Vec-XLS-R\footnote{https://huggingface.co/facebook/wav2vec2-xls-r-300m} \cite{babu2021xls} with AASIST (XLSR-AASIST). It is worth mentioning that we utilize the fifth hidden state from the frozen pre-trained representation as the final feature, in reference to previous research on frozen hidden states \cite{lee22q_interspeech}, which demonstrated the best performance.

\vspace{-8pt}
\subsection{Other Implementation Details}
For pre-processing of the ADD baseline models, all audio samples were first down-sampled to 16,000 Hz and trimmed or padded to a duration of 4 seconds. All models used Adam optimizer with a learning rate of 5$\times10^{-4}$. For the small-scale training sets (19LA, CFAD, FSW), we implemented 50 epochs training with learning rate halving every 10 epochs. For Codecfake training sets, we employed 10 epochs training with 2 epochs interval halving, with the weight for the real class set to 10 and for the fake class set to 1. The model showing the best performance on their corresponding development set was chosen for evaluation.

\vspace{-8pt}
\section{Results and Discussion}
\vspace{-5pt}
\subsection{Countermeasures trained on public datasets}
\vspace{-5pt}
We first conducted baseline training on three architectures (AASIST, WavLM-AASIST, and XLSR-AASIST) using the 19LA, CFAD, and Codecfake training set described in Section \ref{section:public}. As detailed in Table \ref{tab:publicdataset}, our initial analysis reveals that the XLSR-AASIST model trained on 19LA achieves 0.22\% EER on the intra-domain 19LA test set, with the lowest average EER of 11.06\% across five evaluation datasets. However, its performance is significantly compromised by particularly poor results on the ITW (13.58\%) and FSW (32.51\%) datasets, indicating limited generalization capability for wild ADD detection when trained on the 19LA training set. Subsequent experiments employing CFAD and Codecfake training sets for CM development revealed XLSR-AASIST maintained superior performance, achieving average EER of 11.09\% and 8.65\% respectively. Notably, Codecfake-trained XLSR-AASIST demonstrated efficacy in FSW evaluation (21.82\%), suggesting potential prevalence of ALM-based deepfake speech within the FSW dataset. In the final experimental configuration employing tri-training set co-training, the co-trained XLSR-AASIST achieved the lowest ITW, FSW, and average EER of 9.57\%, 17.23\%, and 5.54\%, respectively, which indicates that co-training improved the generalization of the CM and enhanced its performance in real world scenarios.
\vspace{-10pt}
\subsection{Data augmentation experiments}
\label{section:da}
\vspace{-5pt}
In this section, we applied DA to the best CM, co-trained XLSR-AASIST as describe in previous section, to investigate whether DA could further enhance the robustness of CM. The experimental results are shown in Table \ref{tab:da}, which indicate significant improvements on ITW when using MR and RB. Specifically, the CM with MR achieved an EER of 3.58\% on ITW and the lowest average EER of 4.72\% in the DA experiments. For RB, RB4 achieved the best EER of 2.42\% on ITW; however, it showed a noticeable decline in performance on intra-domain such as 19LA. We also considered whether a combination of RB and MR noise augmentations could further enhance CM performance. However, the combination setting did not yield significant improvements on ITW and FSW, and instead, it increased the EER under intra-domain.
\vspace{-7pt}
\subsection{Countermeasures trained on FSW dataset}
\vspace{-5pt}
Due to the FSW training set containing only a small number of accounts and a total of 19,962 audio samples, the CM performance trained on FSW training set was generally poor. However, we were surprised to find that it achieved an average score of 16.40\% on the FSW test set, which may suggest that the limited size of the FSW training set helped the CM adapt to the diverse audio encoding and decoding environments of different platforms, thereby enhancing its robustness across the four platforms.

To achieve optimal performance of the CM across all public datasets as well as the FSW, we conducted joint training using the co-trained training set from previous experiments (19LA, CFAD, Codecfake) along with the FSW training set, resulting in a combined training set of four datasets. The results are shown in the Table \ref{tab:four_training}. The best CM remains XLSR-AASIST. Comparing the best CM trained on co-trained datasets in Table \ref{tab:publicdataset}, we can observe that its performance on the FSW dropping from 17.23\% EER to 12.55\% EER, which led to a decrease in the average EER from 5.54\% with the three-dataset co-trained model to 4.57\%. Furthermore, we applied the best DA strategy, MR, mentioned in Section \ref{section:da}, to augment the relatively clean co-trained dataset. The experimental results indicate that after employing DA strategy, the EER for both ITW and FSW further decreased, with ITW dropping from 9.35\% to 5.24\% and FSW from 12.55\% to 11.58\%. Consequently, the average EER decreased from 4.57\% to 3.54\%. 
\vspace{-8pt}
\section{Conclusion}
\vspace{-6pt}
This paper focuses on developing CMs for detecting deepfake speech on social media platforms. Specifically, we introduce the FSW dataset, which comprises 254 hours of data from four different platforms and multiple accounts. For the CM, we first evaluate the performance of CMs trained on publicly ADD datasets and test on their evaluation set and wild dataset (ITW, FSW). Subsequently, we examine the efficacy of DA strategies applied to these public datasets. Finally, we achieve optimal results by jointly training the DA-enhanced public datasets with the FSW training dataset. Future work will develop corresponding algorithms to bridge the gap between public datasets and real-world ADD scenarios.

\renewcommand{\refname}{\centering References}
\fontsize{8pt}{9.5pt}\selectfont 
\bibliographystyle{IEEEbib} 
\bibliography{myrefs}

\end{document}